\documentstyle[prb,aps]{revtex}
\begin{document}

\title{Predicted giant magnetic moment on non-\{n0m\} surfaces of $d$-wave 
superconductors \\--- Can it be observed and how?}

\author{Chia-Ren Hu$^1$ and Xin-Zhong Yan$^{1,2}$}
  
\address{$^1$Department of Physics, Texas A\&M University, College Station, 
TX 77843-4242 \\
$^2$Institute of Physics, Chinese Academy of Sciences, P. O. Box 603, 
Beijing 100080, China} 

\date{\today}

\maketitle

\begin{abstract}

It has been noted previously that the sizable areal 
density of midgap states which must exist on any non-\{n0m\} 
surface of a $d$-wave superconductor can lead to a giant 
magnetic moment. Here we show that this effect is observable, 
and discuss two precise ways to observe it: (i) by directly 
measuring magnetic moment in a system with a large density 
of internal \{110\} surfaces, or (ii) by performing 
spin-polarized tunneling on a \{110\} surface. In both cases 
a sufficiantly large magnetic field should be applied in the 
[1$\bar 1$0] direction.

\end{abstract}


\twocolumn

One of us (CRH) has noted previously [1] that the sizable areal 
density of ``midgap states'' (MS's) which must exist on any 
non-\{n0m\} surface of a $d$-wave superconductor (DWSC) (with 
$n$ and $m$ integers or zero) can lead to a ``giant magnetic
moment'' (GMM). The MS's are topological signatures of 
unconventional pairing symmetry. They are 
``nearly-dispersionless'' quasi-particle states, 
characterized by momenta along the surface ranging from $-k_F$ 
to $k_F$ ($k_F$ being the Fermi momemtum), but all with  
the same ``zero'' energy as measured from the 
Fermi energy (in the WKBJ approximation).  
These states can lead to a narrow density-of-states (DOS) 
peak at the Fermi energy, where the integrated bulk-DOS  
dips to zero.  One of the observable consequences of 
these MS's is therefore a ``zero bias conductance peak'' 
(ZBCP) in single-particle tunneling [1,2,3], which has been 
observed ubiquitously in high-$T_c$ superconductors (HTSC's)
for more than a decade. (See Ref. [3] for a review.)  
Several carefully-controlled experiments performed 
recently strongly supported the conclusion that the ZBCP's 
observed in them are due to such MS's [4]. 

Many other consequences of the MS's have been predicted 
[5], including a new contribution to Josephson tunneling [6], 
a paramagnetic Meissner effect [7,8], and a 
low-temperature anomaly in the penetration depth [9], 
etc. The GMM is also a consequence of the MS's, 
which has not yet been looked for experimentally, perhaps 
because no detailed analysis has been made on whether and how 
it can be observed.  Thus here we perform such an analysis.

An applied magnetic field $B$ can have two simultaneous 
effects on the surface MS's [1,7,9]: (i) {\it Spin shift:} 
The MS's are spin eigenstates. The field B can cause the 
MS's of one spin to shift above the Fermi surface, and those 
of the other spin to shift below. If the orbital shift 
(described below) can
be neglected, then when the magnitude of the spin shifts 
grows past: (i) the width of the MS's peak in the DOS, 
(ii) the small non-zero energies $\sim\Delta_0^2/E_F$
of the MS's, and (iii) the thermal energy $k_BT$, 
then a measurement of the total magnetic moment of the 
system should exhibit a saturation phenomenon of a magnitude 
proportional to the total number of MS's on the surface. 
This is referred to as GMM in [1].  
(ii) {\it Orbital shift:} (including screeening current effects.) 
In the gauge in which the pair-potential order parameter 
$\Delta$ is real, this effect is due ot a vector potential ${\bf A}$ 
alone. The MS's acquire energy shifts proportional to their momenta 
$k$ along ${\bf A}$. At sufficiently low T all occupied MS's have 
the same sign of $k$, (in the WKBJ approximation and neglecting 
the spin shift,) implying a paramagnetic equilibrium current. 
Higashitani [7] thus proposed that this effect can account 
for the observed ``paramagnetic Meissner effect'' [8]. Another 
predicted consequence is a low temperature anomaly in the ab-plane 
penetration depth $\lambda_{ab}$ which has been observed [9]. 

Both types of energy shifts are really present at the same time. 
Thus it is important to estimate their relative magnitudes. 
We find that the conclusion depends crucially on the direction 
of $B$.  Consider a thick single-crystal slab with a \{110\} 
surface, and with $B$ applied parallel to the surface. If $B$ is 
along [001], the screening current is along [1$\bar 1$0], which 
is denoted as the $y$ axis, with the $x$ axis perpendicular 
to the surface at $x=0$. The spin shift has essentially the 
magnitude $\mu_0B$. ($\mu_0$ is the Bohr magneton.)  
The orbital shift can be estimated using first 
order purturbation theory.  The purtubation Hamiltonian is 
$-(1/c)\hat{\bf j}\cdot {\bf A(x)}$. In the gauge in which 
$\Delta$ is real, 
${\bf A}(x) = -B\lambda_{ab} \exp(-x/\lambda_{ab})\hat e_y 
\simeq {\bf A}(0) = -B\lambda_{ab}\hat e_y$, because the MS's 
are localized within roughly one coherence length 
($\xi_{ab}\sim 15\AA$) from the surface, which is much smaller 
than $\lambda_{ab} \sim 1500\AA$. (For the same reason the spin 
shift sees simply the applied field $B$.)  Thus the orbital 
shift is $\sim (e/m_{ab}c)\vert k\vert B\lambda_{ab}$.  
The ratio between the orbital and spin shifts is  
$2(m_e/m_{ab})(\vert k\vert\lambda_{ab})$. 
The mass ratio $m_e/m_{ab}$ is probably less than unity
by a factor larger than 0.2. [10] $k$ ranges from 
$-k_F$ to $k_F$, and $k_F$ should be somewhat less than 
$\pi/2a$, where $a$ is the lattice parameter in the $ab$ 
plane. Thus this ratio is around 200, showing that with $B$ 
along $[001]$ the spin shift is negligible in comparison 
with the orbital shift, and the previous analyses of the 
consequences of the orbital shift without considering the 
spin shift [7,9] is justified. 

Next let $B$ be along [1$\bar 1$0]. The screening 
current is then along [001].  The same ratio in the 
simplest estimate is now $2(m_e/m_{c})(k_z\lambda_{c})$. 
The London penetration-depth formula implies that 
$m_c\propto \lambda_c^2$. Thus this energy-shift ratio is 
reduced from the previous one by a factor 
$(\lambda_{ab}/\lambda_c)$, which is $\sim 1/50$ for Hg-1201 [11]. 
More careful estimate, taking into account (i) the 
tight-binding nature in the c direction, (ii) $k_z$ ranges 
from $-\pi/c$ to $\pi/c$ with $c/a\sim 3$, reduces the 
energy-shift ratio to ${\buildrel < \over \sim} 2$. We shall see 
that this is quite sufficient to allow the GMM to be observed.  
On the other hand, by forming NS superlattices, it should be 
possible to increase $\lambda_c$ by another factor of 
10 or larger, then one can even explore the regime where the 
orbital shift is negligible.  Below, we first consider the 
effects of a spin shift alone, and then comment later on 
the effects of a simultaneous orbital shift of a comparable 
magnitude.  We shall discuss the spin magnetization first, 
and then the spin polarized tunneling conductance. We believe 
that the latter is a very promising way to see this effect.

Consider first the spin magnetization $M$ (which we deifne as 
the magnetic moment per unit area per CuO$_2$ plane). With
$$M \equiv -\mu_0[<\psi^{\dagger}_{\uparrow}(x)\psi_{\uparrow}(x)> - 
<\psi^{\dagger}_{\downarrow}(x)\psi_{\downarrow}(x)>], \eqno(1)$$
where $\psi_s(x)$ is the field operator of spin-$s$ electrons, 
its main contribution from the MS's follows easily from a 
perturbative treatment of the Bogoliubov-de Gennes (BdG) equations 
[1,3], which gives
$$M(x) = \mu_0\,g(\mu_0B)f(x), \eqno(2)$$
with
$$g(E) = \int^{\infty}_{-\infty}{d\omega\over 2\pi}\,\tanh(\beta\omega/2)
         \,{\rm Im}({1\over\omega+E+i\delta}- 
                                   {1\over\omega-E+i\delta}), \eqno(3)$$
$$f(x) = \int_{-k_F}^{k_F}{dk\over 2\pi}\,
           [\vert u_k(x)\vert^2 + \vert v_k(x)\vert^2], 
                                                              \eqno(4)$$
where $\beta = 1/k_BT$; $u_k(x)$ and $v_k(x)$ are the electron and hole 
components of the wave function of the MS of momentum $k$ along $y$; 
and $\delta > 0$ takes into account a finite life time due 
to surface and/or bulk scatterings. In the limit $\delta \to 0^{+}$,
$g(E)\to\tanh(\beta E/2)$. In general, $g$ is a function of both
$\beta\mu_0B$ and $\beta\delta$, and is a monotonically decreasing 
function of $T$ at any given $B$ and $\delta$. The largest value for 
$g$ is unity, corresponding to $\mu_0B>>\{\delta$ and $k_BT\}$.  
Then the total magnetic moment per CuO$_2$ plane, associated with 
the MS's on one \{110\} edge of the plane, obtained by integrating 
$M(x)$ over $x$ and $y$, is equal to $(2L_y/\lambda_F)\mu_0$. 
That is, for every Fermi wavelength on each $\{110\}$ edge of a CuO$_2$ 
plane, there are two MS electrons contributing to the GMM.  This is 
the maximum magnitude of the saturation phenomenon mentioned earlier.
To observe it directly, however, one needs to drastically increase the 
surface to volume ratio in the sample.  The approach adopted in Ref. [9], 
where a sample is irradiated with ions along [110], in order to 
create a large number of straight tracks, might offer some hope.

Next we consider tunneling between a spin-polarized normal metal 
(N$_{sp}$), ({\it i.e.}, a ferromagnetic metal or a half-metallic 
magnet [12],) and a DWSC (S$_d$) with a \{110\} surface. With $B$ 
along [1$\bar 1$0] and the orbit shift neglected,
the ZBCP should split into two peaks at 
non-zero voltages where the Fermi level of each spin species in 
N$_{sp}$ matches the shifted energy of the MS's in S$_d$ {\it of the 
same spin}. The relative heights of these two peaks should depend 
directly on the spin polarization in N$_{sp}$. Below we 
make these statements more quantitative by considering the effects 
of finite peak width and temperature.

We assume, for simplicity, that both sides of the junction have 
the same carrier type and density, and the same {\it two-dimensional} 
band dispersion (with HTSC's in mind). We also assume that the 
carriers are electrons with charge $-e<0$ and a gyromagnetic ratio 
$\gamma = -ge/2m_ec$ with $g=2$. Later we will comment on the effects 
of replacing these assumptions by more realistic ones, such as a 
three-dimensional band dispersion for N$_{sp}$, and different 
carrier types and densities in the two sides of the junction. 
 
The zero-field polarization in N$_{sp}$ is defined to be:
$$P \equiv (n_{\uparrow} - n_{\downarrow})/(n_{\uparrow} + n_{\downarrow}) 
                                                               \eqno(5)$$
where $n_s$ is the density of spin-$s$ electrons at $B = 0$.  We consider
$P > 0$. The zero-field Fermi energies for spin-$s$ electrons in N$_{sp}$  
can then be expressed as $E_{F,s}^{(0)}/\bar E_F^{(0)} = 1 + sP$ [$s = 1(-1)$ 
for spin $\uparrow$($\downarrow$)], where 
$\bar E_F^{(0)} = (E_{F,\uparrow}^{(0)} + E_{F,\downarrow}^{(0)})/2$, 
and is equal to the Fermi energy $E_F$ in S$_d$ due to our simplifying 
assumptions. The single-particle excitations in this system is governed by 
the BdG equations [13], 
$$\left (\matrix{\hat H_{\uparrow} & \Delta\cr
\Delta^{*} &-\hat H_{\downarrow}}\right )\left ({\matrix {u \cr v}}\right ) 
       = \epsilon \left({\matrix {u \cr v}}\right)            \eqno(6)$$
where $\hat H_s(x)$ is equal to $p^2/2m + s\mu_0B - E_{F,s}^{(0)}$ for $x<0$ 
({\it i.e.}, in N$_{sp}$, valid for $E_{F,\uparrow}^{(0)} > \mu_0B$ only),
and to $p^2/2m + s\mu_0B(x) - E_F$ for $x>0$ ({\it i.e.,} in S$_d$,
$B(x)\simeq B$ for the MS's). 
The pair potential $\Delta$ vanishes in N$_{sp}$, and is 
assumed to be $x$-independent in S$_d$. (Its self-consistency  
needs not be considered, for the MS's are topological.) 
$\epsilon$ is the quasi-particle energy measured from the chemical 
potential, which is $= E_F$ because we take the bottom of the 
conduction band in S$_d$ to be zero. The bottom of the conduction 
band of each spin in N$_{sp}$ is then not zero, and has been
absorbed in the definition of $E_{F,s}^{(0)}$. In momentum space, 
the d-wave pair potential $\Delta({\bf k}_F)$ is taken to be 
$\Delta_0\cos(2\theta-2\alpha)$, where $\theta$ and $\alpha$ are 
the angles ${\bf k}_F$ and the crystal $a$-axis make with the 
x-axis. We consider $\alpha = \pi/4$. At $x$ = 0 a $\delta$-function 
barrier, $H\delta(x)$, is assumed. All wave vectors are 
two-dimensional due to our assumptions. In N$_{sp}$, the two-component 
quasi-particle wave function is, aside from a factor $\exp(iq_yy)$,
$$\Psi_s = \psi_{s,q_x}-a_s\psi_{-s,k_x}-b_s\psi_{s,-q_x}, \eqno(7)$$   
where $\psi_{s,q_x} = \psi_{s}\exp(iq_xx)/\sqrt{|v_{q_x}|}$ 
with $\psi_{\uparrow} = \pmatrix{1\cr 0}$ for a spin-up electron, 
and $\psi_{\downarrow} = \pmatrix{0\cr 1}$ for a spin-down hole, 
and $v_{q_x} = d\epsilon/dq_x$ is the qroup velocity. The 
first term in the right hand side of Eq. (7) expresses the 
incoming quasiparticle with spin-$s$ and wave number $q_x$ 
normal to the interface.  The other two terms correspond to Andreev 
and ordinary reflections, respectively, with $a_s$ and $b_s$ their 
respective coefficients.  Eq. (6) gives 
$\hbar^2q^2_x/2m = E_{F,s}^{(0)} -s\mu_0B + s\epsilon - \hbar^2q^2_y/2m$, 
and $\hbar^2k^2_x/2m = E_{F,-s} +s\mu_0B - s\epsilon - \hbar^2q^2_y/2m$.  
The transmitted wave in S$_{d}$ is a linear combination of 
outgoing waves (from the interface) that are solutions of the 
BdG equations for a bulk DWSC. Matching the wave function at 
the interface gives $a_s$, $b_s$, and other coefficients in 
the transmitted waves.
 
The tunneling conductance $G$, (normalized to unity at 
$E_F >> eV >> \Delta_0$ when $T$, $P$, and $B$ are all 0,) 
is calculated using the Blonder-Tinkham-Klapwijk formulism: [14,15]   
$$\begin{array}[b]{rl}
G &= -\frac{\hbar^2}{4q_F}\int_0^\infty qdq \int_{-\pi/2}^{\pi/2}d\phi 
     \frac{q\cos\phi}{m}\{\sum_{s=\uparrow,\downarrow}\cr\cr
&f'(s\epsilon+eV)[1+A_{s}(\epsilon, \phi)-B_{s}(\epsilon, \phi)]\},
\end{array}\eqno(8)$$
where $\phi$ is the angle between ${\bf q}$ and the $x$ axis, 
$\epsilon(s,q,\phi)$ is given in the previous paragraph, 
$q_F\equiv (2mE_F)^{1/2}/\hbar$, 
$f(\epsilon)\equiv 1/[\exp(\epsilon/k_BT)+1]$,
and $A_s\equiv |a_s|^2$, $B_s\equiv |b_s|^2.$ 

For numerical calculation, we take $\Delta_0/E_F = 0.08$ as a typical 
value for HTSC's. The dimensionless barrier parameter is 
$Z = {H\over \hbar}\sqrt{m\over 2E_F}$ [14].  Fig.~1 gives $G$ as a 
function of $V$ at $Z$ = 5, for $B = T = 0$ and for
$\mu_0B = 0.03\Delta_0$, $k_BT = (0,\mbox{ }0.2, \mbox{ and }0.4)\mu_0B$, 
for a non-magnetic N$_{sp}$, $P = 0$.  
At $B = 0$, we get the ZBCP as observed in many experiments. At  $B \ne 0$, 
the conductance peak splits into two peaks at $eV = \pm\mu_0B$ which 
correspond to the energy levels of the surface states of different spins 
in the SC. 

At $\mu_0B>>(k_BT$ and the peaks' width), the spin-down surface states with 
energy $\epsilon = -\mu_0B$ are occupied with electrons, whereas the spin-up 
surface states at energy $+\mu_0B$ are empty. As a voltage $V > 0$ is
applied between N$_{sp}$ and S$_d$, the chemical potential in N$_{sp}$ is 
lowered by $eV$. The spin-down electrons in the surface states in S$_{d}$
will tunnel into N$_{sp}$ only when $eV$ increases past $\mu_0B$, leading 
to a step-like increase in the tunneling current, or a conductance peak. If 
a negative voltage $V$ is applied between N$_{sp}$ and S$_d$, the chemical 
potential in N$_{sp}$ is increased by $e\vert V\vert$. When it exceeds 
$\mu_0B$, The spin-up electrons in N$_{sp}$ can then tunnel into the 
empty spin-up MS's in S$_d$, leading also to a peak in $G$.  

Figure 2 gives a similar plot for $G$ at $P$ = 0.5 (without the  
$B = 0$ case, which does not change with $P$). The two conductance 
peaks at $eV = \pm\mu_0B$ now have different heights.  The peak 
associated with tunneling of spin-down electrons is lower, because 
their Fermi velocity in N$_{sp}$ is smaller. (The DOS in two dimensions 
is a constant of energy, otherwise there would be another source for 
the peak-height difference.)
 
Figure 3 gives $G$ when N$_{sp}$ is fully polarized, $P$ = 1, as 
found in some half metallic magnets [12]. In this case only one 
peak appears, which is associated with the tunneling of spin-up 
electrons. 

The absolute heights of these peaks have meaning, as $G$ has been 
normalized.  Note that the conductance peaks are higher for 
larger $Z$. At the same time, they become narrower.  This is because, 
for higher barrier, the lifetimes of the surface states in 
S$_d$ become longer, and there is a sharper resonance between 
the particles from N$_{sp}$ and the MS's in S$_d$.  Therefore, to 
detect the GMM this way, it is preferable to work in the ``tunneling 
limit'', {\it i.e.,} when the interfacial barrier is high. 

A typical value for $\Delta_0$ in HTSC's is about 16.5 meV.  
Then to reach the energy $0.03\Delta_0$, the magnetic field needs 
to be around 8.6 Tesla. In addition, $0.03\Delta_0$ in temperature 
is about 5.7$K$. To measure such an energy shift, the experiment 
should be performed at liquid helium or lower temperatures.  

If an orbital shift of the same order as the spin shift is present in
the system, then each peak will become wider and approximately
rectangularly shaped. But the effects predicted here are still observable, 
even if the maximum orbital shift is, say, three times larger than the spin 
shift. Correction to the WKBJ approximation has been neglected, otherwise
$\mu_0B$ might have to be larger to see this effect. (Accurate estimate of this correction is difficult, but it is of the order $\Delta_0^2/E_F$.)

If the carrier density in N$_{sp}$ is different from that in S$_d$, or if 
the band dispersion relations of the two sides are different, then the main 
effects, as far as we can see, are: (i) a change in the absolute magnitude of 
the tunneling conductance which does not affect the normalized conductance,  
(ii) a stronger $P$ dependence due to the energy dependence of the DOS 
in N$_{sp}$ (without changing the limiting behavior at $P=1$), and (iii)
a change in the ordinary reflection coefficient at the interface,
which can be simulated with an effective $Z$. Even when the carriers 
in S$_d$ are holes, and those in N$_{sp}$ are electrons, we still find 
no essential change in our predictions. 

In summary, we have analyzed in detail the giant magnetic moment that 
can result from the midgap states on the \{110\} surface of a $d$-wave
superconductor. With high-$T_c$ superconductors most-likely having 
$d$-wave pairing, this predicted giant magnetic moment should be 
observable either directly in samples with a high concentration of 
internal \{110\} surfaces generated by ion irradiation, or, with better 
promise, by measuring the tunneling conductance between a spin-polarized 
normal metal and a high-$T_c$ superconductor with a \{110\} surface. 
At a large enough external magnetic field applied along [1$\bar 1$0], 
and low enough temperature, the ZBCP is shown to split into two peaks, 
with their relative heights determined by the polarization of the normal 
metal, because each peak is associated with a single spin. Observing
these predictions can confirm the existence of the midgap states, 
unconventional pairing, and remove any remaining doubt that the observed 
zero-bias conductance peak might not be due to the midgap states.

This work is supported by the Texas Higher Education Coordinating 
Board under grant \# 1997-010366-029, and by the Texas Center 
for Superconductivity at the University of Houston.


Figure Captions
\begin{itemize}


\item[Fig. 1]
Normalized tunneling conductance $G$ between an unpolarized ($P=0$) normal metal 
and a $d$-wave superconductor with a \{110\} surface, as a function of voltage $V$, 
for $B = 0$, $T = 0$, and $B = 0.03\Delta_0/\mu_0$ at three values of $T$. 
The interfacial barrier parameter $Z = 5$. Only the contributions from the 
midgap states are included. 

\item[Fig. 2]
Similar plot for $G$ except that the normal metal is 50\% spin-polarized ($P = 0.5$). 
Only the $B\ne 0$ case is plotted since the $B=0$ case is practically independent 
of $P$.
 
\item[Fig. 3]
Same as Fig.~2 except that the normal metal is 100\% spin-polarized ($P = 1$). 

\end{itemize}
 

\begin{thebibliography}{20}

\bibitem{[1]}
C.-R. Hu, Phys. Rev. Lett. {\bf 72}, 1526 (1994); J. Yang and C.-R. Hu,
Phys. Rev. B{\bf 50}, 16\,766 (1994).

\bibitem{[2]}
Y. Tanaka and S. Kashiwaya, Phys. Rev. Lett. {\bf 74}, 3451 (1995); 
S. Kashiwaya {\it et al.}, Phys. Rev. B {\bf 53}, 2667 (1996);
J.-H. Xu {\it et al.}, Phys. Rev. B{\bf 53}, 3604 (1996).

\bibitem{[3]}
C.-R. Hu, Phys. Rev. B {\bf 57}, 1266 (1998). 

\bibitem{[4]} M. Covington {\it et al.} Phys. Rev. Lett. {\bf 79}, 772
(1997); L. Alff {\it et al.}, Phys. Rev. B{\bf 55}, 14\,757 (1997);
S. Sinha and K.-W. Ng, Phys. Rev. Lett. {\bf 80}, 1296 (1998);
J. Y. T. Wei {\it et al.}, Phys. Rev. Lett. {\bf 81}, 2542 (1998).

\bibitem{[5]} 
For a review, see C.-R. Hu, to be published in J. Mod. Phys. B 
(1998). 

\bibitem{[6]} 
Yu. S. Barash, H. Burkhardt, and D. Rainer, Phys. Rev. Lett. {\bf 77},
4060 (1996); M. P. Samanta and datta, Phys. Rev. B {\bf 55}, R8689 (1997). 

\bibitem{[7]}
S. Higashitani, J. Phys. Soc. Jpn. {\bf 66}, 2556 (1997). 

\bibitem{[8]}
W. Braunisch {\it et al.}, Phys. Rev. Lett. {\bf 68}, 1908 (1992).

\bibitem{[9]}
H. Walter {\it et al.}, Phys. Rev. Lett. {\bf 80}, 3598 (1998).

\bibitem{[10]}
A. Ino {\it et al.}, Phys. Rev. B {\bf 81}, 2124 (1998). This reference
gives $m_{ab}/M_b\simeq 1 - 3$ for La$_{2-x}$Sr$_x$CuO$_4$, where $M_b$ 
is the band mass, which we estimate to be comparable to the free electron 
mass $m_e$ based on L. F. Mattheiss, Phys. Rev. Lett. {\bf 58}, 1028 
(1987).

\bibitem{[11]} 
J. R. Kirtley {\it et al.}, Phys.Rev. Lett. {\bf 81}, 2140 (1998).

\bibitem{[12]} J. H. Park {\it et al.}, Nature {\bf 392}, 794 (1998);
C. T. Tanaka and J. S. Moodera, J. Appl. Phys. {\bf 79}, 6265 (1996).

\bibitem{[13]}
M. J. M. de Jong and C. W. J. Beenakker, Phys. Rev. Lett. {\bf 74}, 1657 
(1995).

\bibitem{[14]}
G. E. blonder, M. Tinkham, and T. M. Klapwijk, Phys. Rev. B {\bf 25}, 4515 
(1982).

\bibitem{[15]}
I. \v Zuti\'c and O. T. Valls, to be published. 


\end{thebibliography}
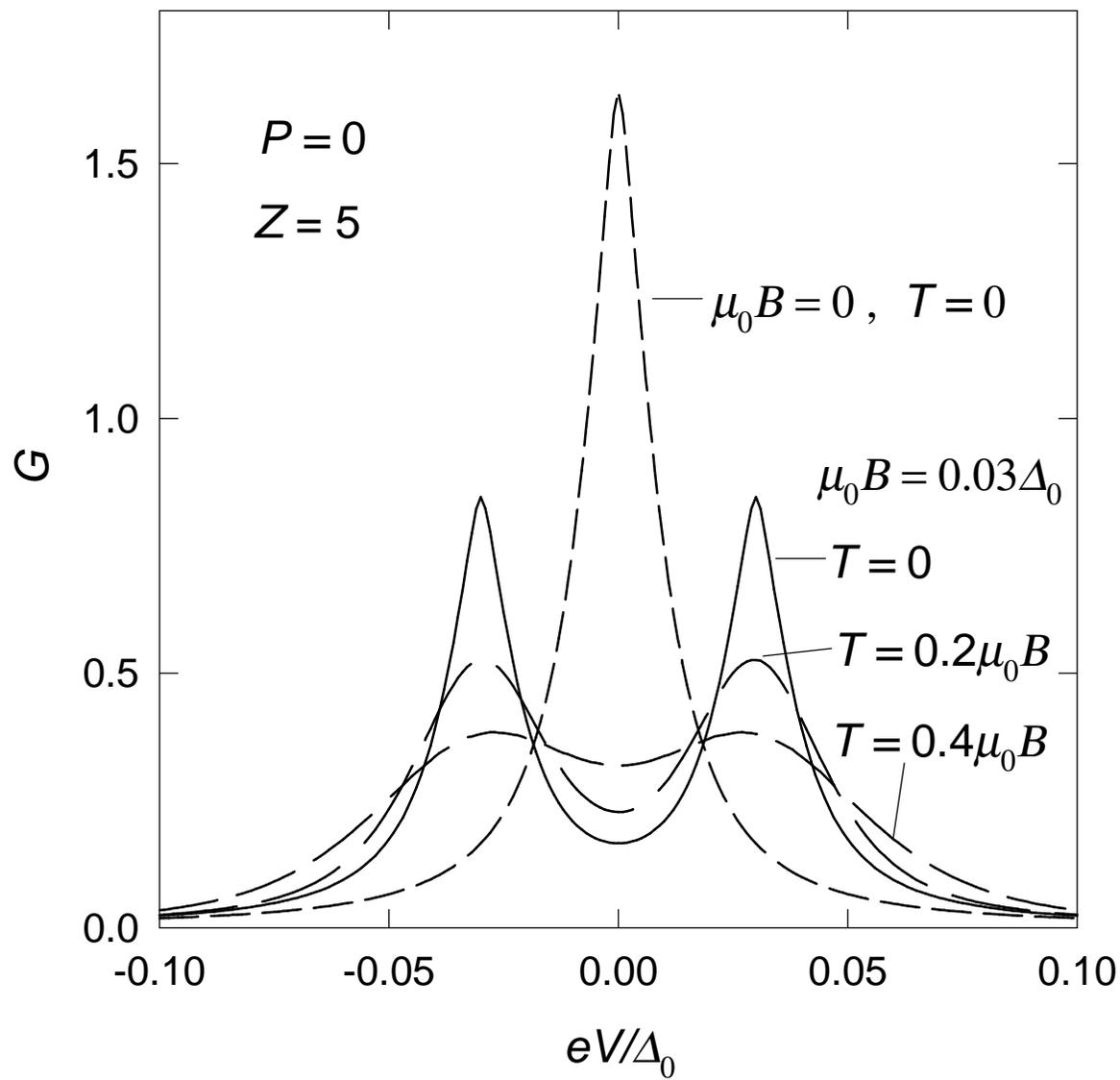
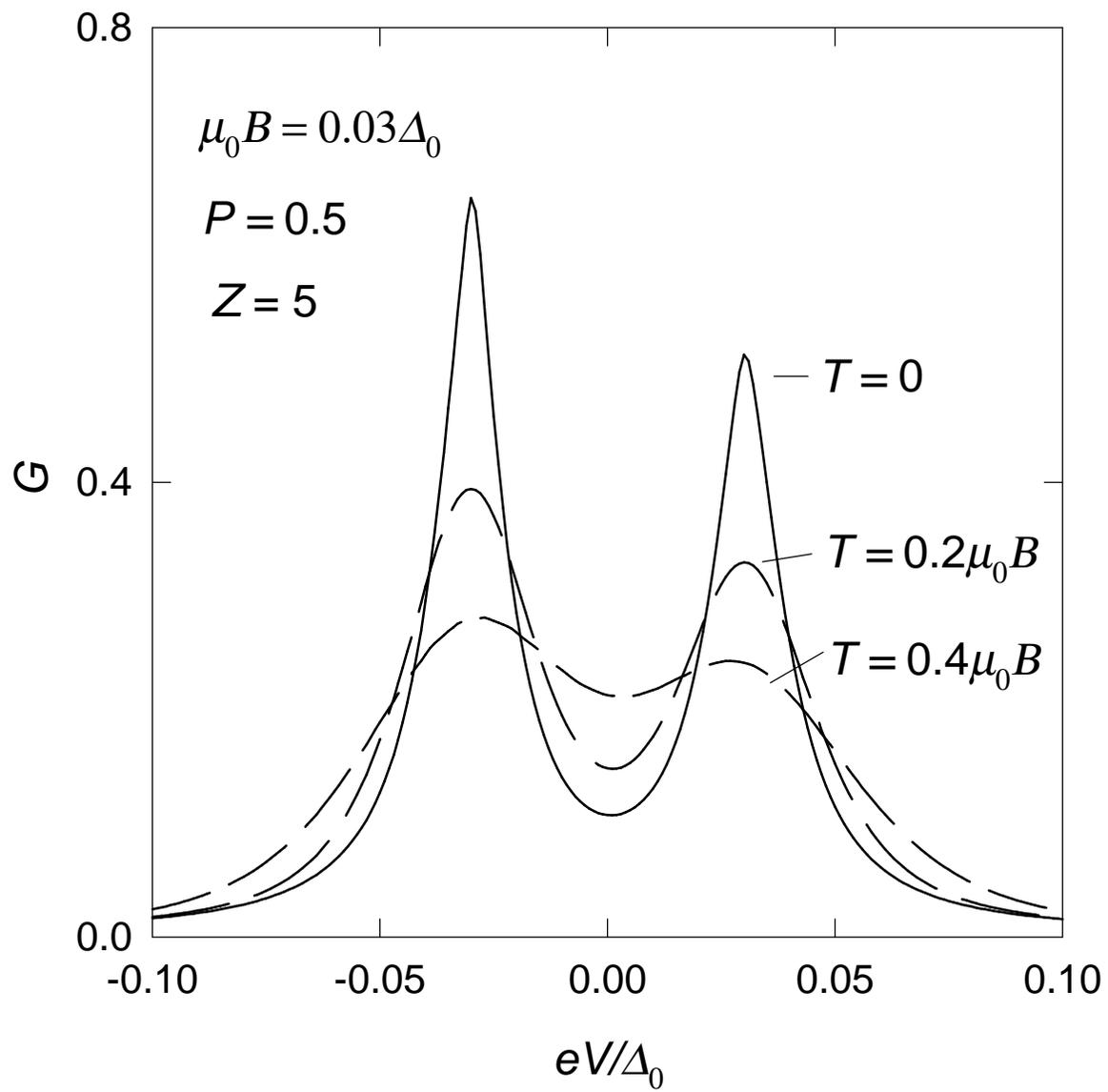
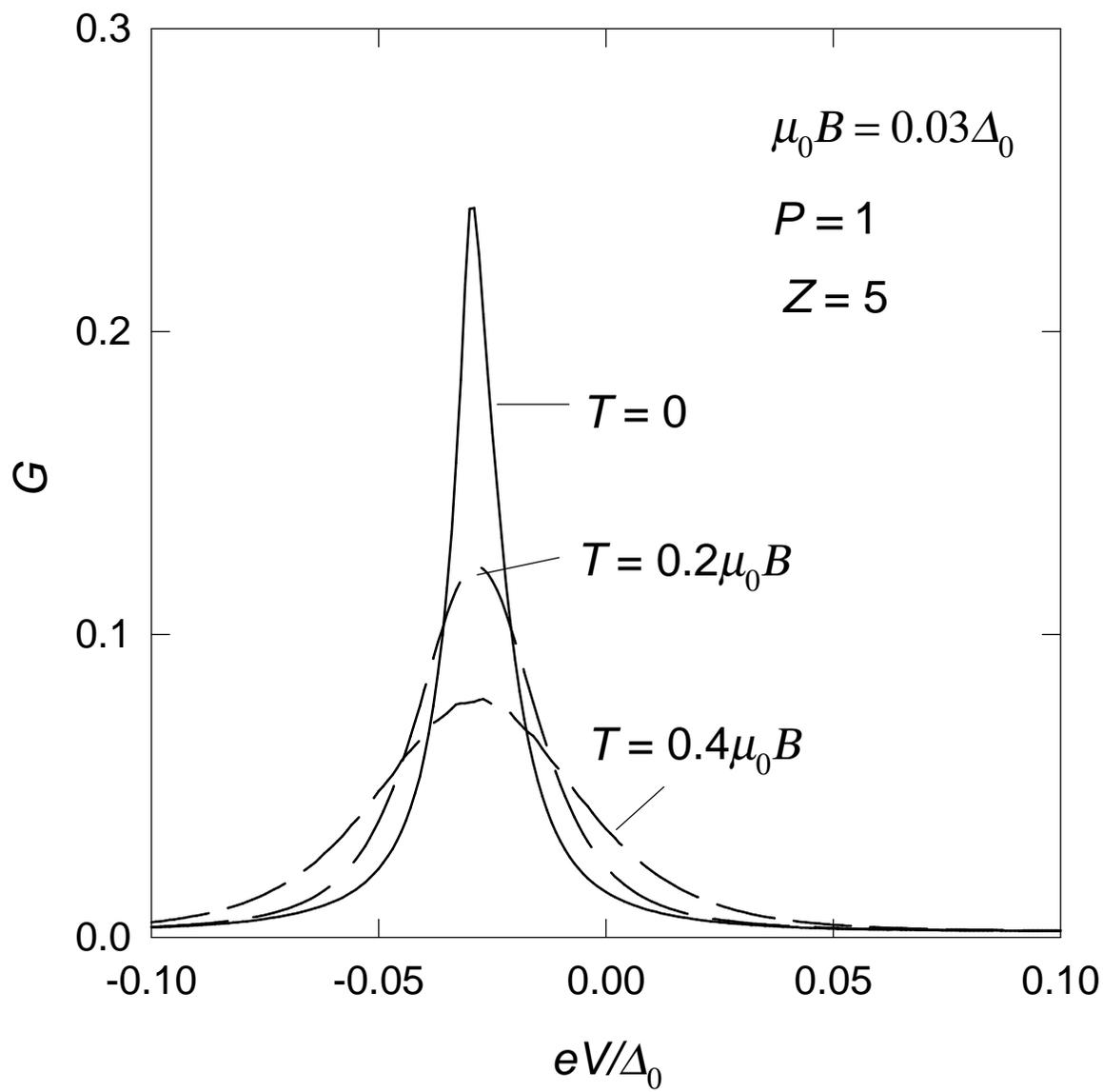
\end{document}